\documentclass[apj]{emulateapj}
\usepackage{times}
\usepackage[normalem]{ulem}
\usepackage{epsfig}
\usepackage{natbib}
\usepackage{amsfonts}
\usepackage{amsmath}
\usepackage{multirow}
\usepackage{enumerate}
\usepackage{verbatim}
\usepackage[usenames]{color}
\usepackage[plainpages=false, colorlinks=true, anchorcolor=blue, linkcolor=blue, citecolor=blue, bookmarks=false]{hyperref}
\newcommand{\be}{\begin{eqnarray}}
\newcommand{\ee}{\end{eqnarray}}

\newcommand{\lp}{\left(}
\newcommand{\rp}{\right)}

\newcommand{\slugcom}{Accepted for publication in The Astrophysical Journal Letters}
\slugcomment{\slugcom}

\begin{document}

\normalsize

% -----------------------------------------------------------
% -----------------------------------------------------------

\title{The Impact of a Supernova Remnant on Fast Radio Bursts}

\author{Anthony L. Piro}

\affil{Carnegie Observatories, 813 Santa Barbara Street, Pasadena, CA 91101, USA; piro@obs.carnegiescience.edu}

\begin{abstract}
Fast radio bursts (FRBs) are millisecond bursts of radio radiation whose progenitors so far remain mysterious. Nevertheless, the timescales and energetics of the events have lead to many theories associating FRBs with young neutron stars. Motivated by this, I explore the interaction of FRBs with young supernova remnants (SNRs), and I discuss the potential observational consequences and constraints of such a scenario. As the SN ejecta plows into the interstellar medium (ISM), a reverse shock is generated that passes back through the material and ionizes it. This leads to a dispersion measure (DM) associated with the SNR as well as a time derivative for DM. Times when DM is high are generally overshadowed by free-free absorption, which, depending on the mass of the ejecta and the density of the ISM, may be probed at frequencies of $400\,{\rm MHz}$ to $1.4\,{\rm GHz}$ on timescales of $\sim100-500\,{\rm yrs}$ after the SN. Magnetic fields generated at the reverse shock may be high enough to explain Faraday rotation that has been measured for one FRB. If FRBs are powered by the spin energy of a young NS (rather than magnetic energy), the NS must have a magnetic field $\lesssim10^{11}-10^{12}\,{\rm G}$ to ensure that it does not spin down too quickly while the SNR is still optically thick at radio frequencies. In the future, once there are distance measurements to FRBs and their energetics are better understood, the spin of the NS can also be constrained.
\end{abstract}


\section{Introduction}

Fast radio bursts (FRBs) are millisecond bursts of radio radiation that have been discovered in pulsar surveys \citep{Lorimer07,Keane12,Thornton13,Ravi15}.
They may be occurring at cosmological distances \citep[see discussions in][and references therein]{Kulkarni14,Luan14,Katz16}, and have been inferred to happen at an incredible rate of $\sim10^4$ FRBs on the sky per day \citep{Rane16}. Thus far, there has been no astrophysical object or event definitively connected to FRBs, which has inspired a large number of theoretical studies to solve the mystery of identifying their progenitor (perhaps more than the total number of FRBs now detected). This includes neutron stars (NSs) collapsing to black holes \citep[BHs,][]{Falcke14}, asteroids and comets falling onto NSs \citep{Geng15,Dai16}, giant pulses or bursts from various age and magnetic field strength NSs \citep{Cordes16,Connor16,Lyutikov16}, circumnuclear magnetars \citep{Pen15}, flaring stars \citep{Loeb14}, merging charged BHs \citep{Zhang16}, white dwarf mergers \citep{Kashiyama13}, and magnetic NS mergers \citep{Hansen01,Piro12,Wang16}.

One of the most constraining properties of FRBs is that at least one is repetitive with 17 bursts over almost 3 years \citep{Spitler14,Spitler16,Scholz16}. It is still not clear whether other FRBs repeat like this, but if they do it would be difficult to reconcile with any catastrophic scenario. This has inspired a range of discussions on whether FRBs might be related to soft gamma-ray repeaters \citep[SGRs,][]{Kulkarni14,Kulkarni15,Lyubarsky14} or giant pulses from young pulsars \citep{Katz15,Cordes16,Lyutikov16}. In such cases, the supernova remnant (SNR) that was generated in the event that made this NS may still be present, and it may affect the ability of radio waves to propagate out of the system.

Motivated by this, here I study the impact of SNRs on FRBs generated within their interiors. This was also previously discussed by \citet{Connor16} and \citep{Lyutikov16}, but without a detailed treatment of the SNR evolution that is crucial for FRB propagation. In Section \ref{sec:remnant}, I describe the general properties of the SNR and then use this to calculate the interaction of SNRs with FRBs. In Section \ref{sec:constraints}, I revisit young NS models in which the spin energy is meant to power the FRB, and I discuss the additional constraints places by the conclusions here. In Section \ref{sec:discussion}, I summarize my main results and discuss potential future work.

\section{Supernova Remnant Evolution and Impact}
\label{sec:remnant}

As an SN expands and cools, the material recombines over the timescale of $\sim\,{\rm months}$ to a $\sim\,{\rm year}$. This would allow radio emission to freely propagate from an FRB-producing NS down in the center of the ejecta, but unfortunately this situation does not last. The interaction of the SNR with the interstellar medium (ISM) creates a reverse shock that passes back through the ejecta. This shock reaches temperatures sufficient to reionize the material, producing free electrons that can now once again disperse radio emission. The key is to understand what timescales this should be occurring over with respect to when FRBs are expected to be generated.

\subsection{Supernova Remnant Properties}

\begin{figure}
\epsscale{1.2}
\plotone{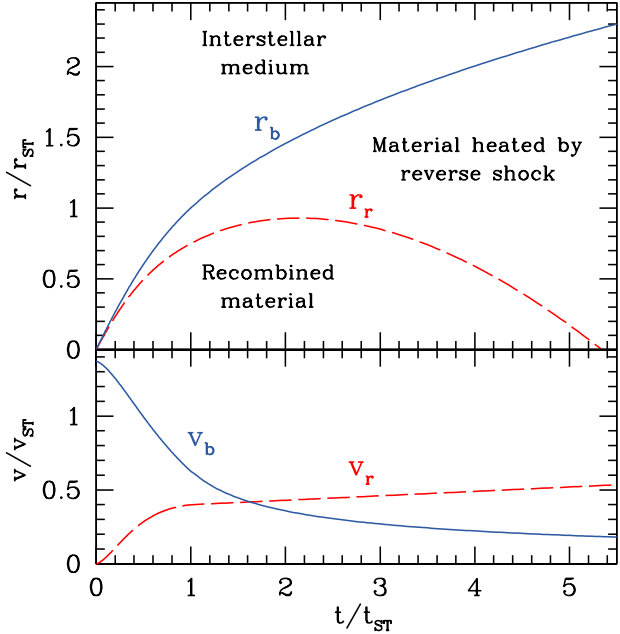}
\caption{The evolution of an SNR using the analytic solutions from \citet{McKee95}. The values $t_{\rm ST}$, $r_{\rm ST}$, and $v_{\rm ST}$ summarize the main properties of the Sedov-Taylor phase, as described in the text below. In the top panel, the radii of the blastwave $r_b$ and reverse shock $r_r$ are plotted. The general picture is that material between these two radii has been shock heated sufficiently for ionization and contributes to dispersing radio waves. The bottom panel summarizes the blastwave shock velocity $v_b$, and the velocity of the reverse shock in the rest frame of the unshocked ejecta just ahead of it $v_r$.}
\label{fig:snr}
\epsscale{1.0}
\end{figure}

To investigate this general picture, I make use of the analytic models presented in \citet{McKee95}. Here I summarize the main features that are relevant for this work. Consider an explosion with energy $E$ and ejecta mass $M_{\rm ej}$. The remnant goes through two main stages: (1) the ejecta dominated stage where the ejecta is still moving out at roughly the velocity set by the SN, and (2) the Sedov-Taylor stage when the ejecta begins to slow from interaction with the ISM. The latter begins when the swept up mass is roughly 1.5 times the ejecta mass and on a timescale
\be
	t_{\rm ST} = 1.4\times10^3E_{51}^{-1/2}M_{10}^{5/6}n_0^{-1/3}\,{\rm yr},
\ee
where $E_{51}=E/10^{51}\,{\rm erg}$, $M_{10}=M_{\rm ej}/10\,M_\odot$, and $n_0$ is the number density of the ISM. Also associated with this is a characteristic lengthscale,
\be
	r_{\rm ST} = 4.8M_{10}^{1/3}n_0^{-1/3}\,{\rm pc},
\ee
and velocity,
\be
	v_{\rm ST} = r_{\rm ST}/t_{\rm TS} =  3.3\times10^3E_{51}^{1/2}M_{10}^{-1/2}\,{\rm km\,s^{-1}}.
\ee
The main features of the analytic solutions are summarized in Figure \ref{fig:snr}. Here $r_b$ represents the radius of the blastwave radius, $r_r$ is the radius of the reverse shock, $v_b$ is the blastwave shock velocity, and $v_r$ is the velocity of the reverse shock in the rest frame of the unshocked ejecta just ahead of it. Given these dimensionless solutions, the properties of the SNR can be rescaled for consideration of any specific SN scenario.

\begin{figure}
\epsscale{1.2}
\plotone{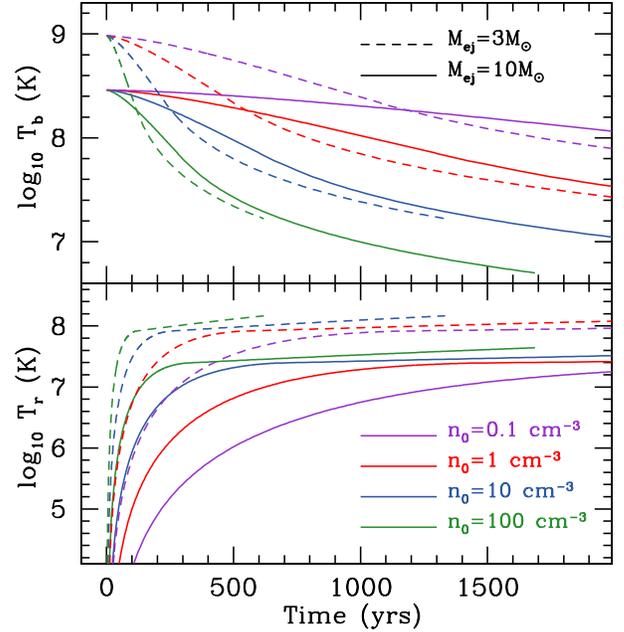}
\caption{Postshock temperatures for the blastwave $T_b$ (upper panel) and the reverse shock $T_r$ (bottom panel) for $M_{\rm ej}=3\,M_\odot$ (dashed lines) and $10\,M_\odot$ (solid lines). The colors indicate the density of the ISM with $n_0=0.1\,{\rm cm^{-3}}$ (purple), $1\,{\rm cm^{-3}}$ (red), $10\,{\rm cm^{-3}}$ (blue), and $100\,{\rm cm^{-3}}$ (green). Lines are only plotted up to the end of the Sedov-Taylor phase, which occurs at a time $\approx5.3t_{\rm ST}$.}
\label{fig:temperature}
\epsscale{1.0}
\end{figure}

One of the key aspects of the ejecta with respect to propagation of radio waves is its temperature. The characteristic scale of the temperature of the shocked gas is
\be
	T_{\rm TS} = \frac{3\mu m_p}{16k_{\rm B}}v_{\rm ST}^2 = 1.5\times10^8E_{51}M_{10}^{-1}\,{\rm K},
\ee
where $\mu$ is the mean molecular weight, $m_p$ is the proton mass, and $k_{\rm B}$ is Boltzmann's constant. This generally shows that the temperatures are sufficient to ionize the gas. The postshock temperatures for the blastwave and reverse shock are
\be
	T_b = (v_b/v_{\rm ST})^2T_{\rm ST},
\ee
and
\be
	T_r = (v_r/v_{\rm ST})^2T_{\rm ST},
\ee
respectively. These are plotted in Figure \ref{fig:temperature} for a range of $M_{\rm ej}$ and $n_0$. The two mass choices of $3\,M_\odot$ and $10\,M_\odot$ are meant to roughly represent stripped and non-stripped SNe, respectively. Generally speaking, a smaller $n_0$ results in a cooler reverse shock temperature. A larger $M_{\rm ej}$ results in smaller temperatures overall.

\begin{figure}
\epsscale{1.2}
\plotone{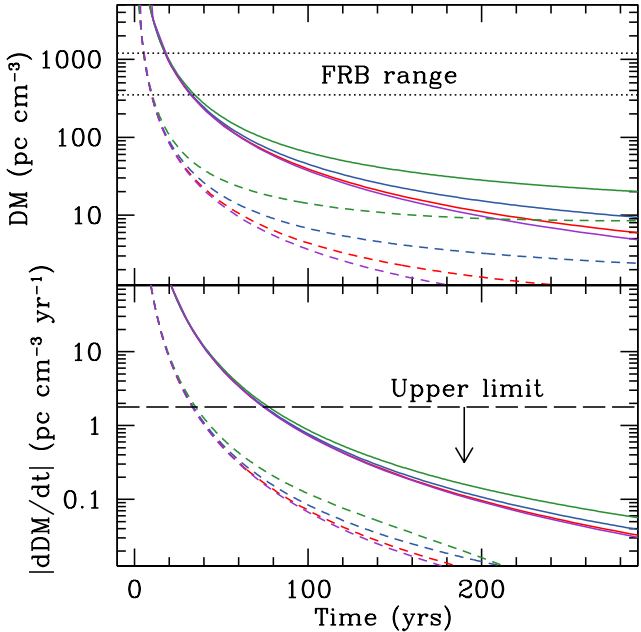}
\caption{The DM (upper panel) and its derivative $d{\rm DM}/dt$ (bottom panel) for the same set of models plotted in Figure \ref{fig:temperature}. In the upper panel, I also include the range of DM values typically seen for FRBs of $350-1200\,{\rm pc\,cm^{-3}}$ (black, dotted lines). In the bottom panel, I note limits on the change in the DM of the repeating FRB 121102, which I estimate as $\lesssim2\,{\rm pc\,cm^{-3}\,yr^{-1}}$ (black, dashed line). }
\label{fig:dm}
\epsscale{1.0}
\end{figure}

\subsection{Dispersion Measure and Faraday Rotation}

With this general picture in mind, one can now estimate how FRBs will be impacted by the remnant. The first consideration is the dispersion measure (DM) added by the SNR. If I only consider regions that have been reionized by the reverse shock, then
\be
	{\rm DM} = n_e\Delta r,
\ee
where $n_e=3M_{\rm ej}/(4\pi r_b^3\mu_em_p)$ is the electron number density, with $\mu_e$ the mean molecular weight per electron, and $\Delta r=r_b-r_r$.  This is summarized in the upper panel of Figure \ref{fig:dm} for a range of models. This generally shows that $n_0$ plays less of a role than $M_{\rm ej}$ on the timescales where the DM is large since $t\lesssim t_{\rm ST}$. Nevertheless, the SNR can contribute $\sim10\%$ of the DM on a timescale of $\sim100\,{\rm yrs}$. The bottom panel of Figure \ref{fig:dm} shows the derivative $d{\rm DM}/dt$. Detection of a decreasing DM may be a way to infer the presence of material local to the FRB even when the SNR DM does not dominate. As a comparison, I consider the repeating burst FRB 121102. Given the error estimate for the DM measured over the few years this burst repeated \citep{Spitler14,Spitler16,Scholz16}, I estimate that the derivative is probably $\lesssim2\,{\rm pc\,cm^{-3}\,yr^{-1}}$.

\begin{figure}
\epsscale{1.2}
\plotone{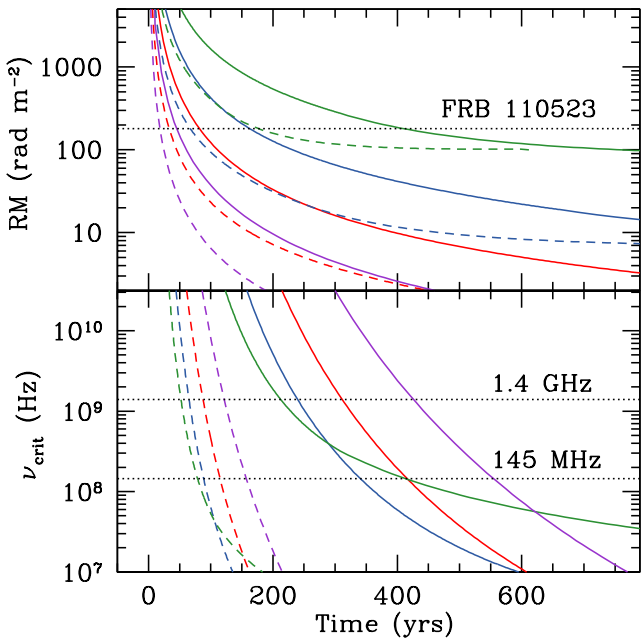}
\caption{The upper panel plots the RM calculated using Equation (\ref{eq:rm}) and a magnetic field set to equipartition with $\epsilon_B=0.1$. The models are the same as in Figure \ref{fig:temperature}. This shows that the Faraday rotation seen for at least one FRB could be explained from the magnetic field in the SNR. In the lower panel I plot the critical frequency $\nu_{\rm crit}$, below which radio emission cannot escape the SNR. Horizontal dotted lines indicate a range of frequencies that radio transient searches will probe.}
\label{fig:freefree}
\epsscale{1.0}
\end{figure}

Magnetic fields generated within the SNR can also have an observable impact on the FRB by generating Faraday rotation. In fact, \citet{Masui15} find a rotation measure ${\rm RM}=-186.1\,{\rm rad\,m^{-2}}$ for one FRB. This is given by
\be
	{\rm RM} = \frac{e^3B}{2\pi m_e^2 c^4}n_e\Delta r,
	\label{eq:rm}
\ee
where $B$ is the magnetic field the radiation is passing through. To set the magnetic field, I assume that it is roughly in equipartition with the reverse shock. This results in $B\approx (4\pi\epsilon_B \rho v_r^2)^{1/2}$, where $\rho$ is the density and $\epsilon_B$ is a parameter that sets how much of the shock energy goes into the magnetic field.  This produces typical field strengths in the range of $B\sim10^{-4}-10^{-3}\,{\rm G}$. The orientation and coherence of the magnetic field can also impact the strength of the RM, and this should be compared with more detailed calculations. Here I absorb this uncertainty into the parameter $\epsilon_B$.

The resulting values of RM are plotted in the upper panel of Figure \ref{fig:freefree}. As a comparison, the RM from FRB 110523 is also indicated \citep{Masui15}.  This shows that this RM could be reasonably explained by SNR magnetic fields. Note that here the electrons producing the RM are likely distinct from the electrons producing the large observed DM because on timescales of $\gtrsim100\,{\rm yrs}$ the DM of the SNR is too small (unlike in the discussions by \citealp{Connor16}). A large range of other values for RM are also possible, which will hopefully be explored by future FRB observations.

\subsection{Free-free Absorption}

An especially important way in which the SNR can impact the FRB is through free-free absorption. For radiation at frequencies $h\nu\ll k_{\rm B}T_r$, the absorption coefficient is \citep{Rybicki79}
\be
	\alpha_{\rm ff} = 1.9\times10^{-2}T_r^{-3/2} Z^2n_en_i\nu^{-2}g_{\rm ff}\,{\rm cm^{-1}},
\ee
where $Z$ is the average charge per ion, $n_i$ is the ion number density, $g_{\rm ff}\sim1$ is the Gaunt factor, and all quantities are in cgs units. Note I use the reverse shock temperature $T_r$, since this will be setting the temperature for most of the ionized SNR material. Setting $\alpha \Delta r=1$, I solve for the critical frequency $\nu_{\rm crit}$, below which radio radiation will not escape.

The results for $\nu_{\rm crit}$ are summarized in the lower panel of Figure \ref{fig:freefree}. This shows that free-free absorption can dominate on timescales of $\sim100-500\,{\rm yrs}$. This means that in most cases the DM associated with the supernova remnant cannot be the observed DM as suggested by \citet{Connor16}. On the other hand, comparing the upper and lower panels of Figure~\ref{fig:freefree}, RM can still be appreciable when free-free absorption is weak. Again $M_{\rm ej}$ plays a large role in setting when the emission can escape, but $n_0$ also has some impact. For smaller $n_0$, the reverse shock is weaker, which in turn causes the temperature to be smaller with more associated free-free absorption. This can alter the time when radio emission can leave by hundreds of years. If this process explains why low frequency observations have not detected FRBs \citep{Karastergiou15,Rowlinson16}, it would imply NSs younger than $\approx600\,{\rm yrs}$ old (or even younger depending on the amount of ejecta mass).

\section{Constraints on Neutron Star Parameters}
\label{sec:constraints}

The calculations in the previous section constrain the NS producing FRBs to be greater than $\sim50-100\,{\rm yrs}$ old at the least and potentially $\gtrsim500\,{\rm yrs}$ if the ejecta mass is especially large and the ISM density small. What does this imply for the properties of this NS? It is difficult to address the radio emission directly, since even for normal pulsars the radio emission has not been derived from first principles. Nevertheless, basic energy and timescale arguments can still be applied \citep[e.g.,][]{Lyutikov16,Katz16}.

\begin{figure}
\epsscale{1.2}
\plotone{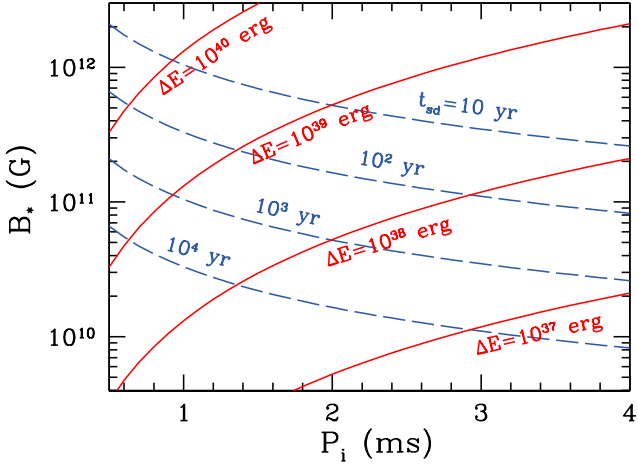}
\caption{Constraints on the possible NS magnetic field $B_*$ and initial spin period $P_i$ to produce FRBs powered by NS spin energy. The red solid curves are lines of constant $\Delta E$, where I have used Equation~(\ref{eq:deltae}) with \mbox{$\Delta t=10^{-3}\,{\rm s}$}. This is the energy required to make an FRB, which could be larger than the isotropically inferred energy if the radio is just a small fraction of the energy release, or could alternatively be smaller than the isotropically inferred energy if the FRB is strongly beamed. The blue, dashed curves are lines of constant spin down time $t_{\rm sd}$. For example, the constraint implied if free-free absorption requires $t_{\rm sd}\gtrsim10^2\,{\rm yrs}$  is that $B_*\lesssim10^{11}-10^{12}\,{\rm G}$.}
\label{fig:constraints}
\epsscale{1.0}
\end{figure}

Here I consider what constraints are placed if the FRB is powered by an NS's spin energy. Consider an NS spinning down due to its magnetic dipole. The amount of energy available in a time $\Delta t$ is
\be
	\Delta E(t)  = \frac{(B_*R_*^3)^2\Omega_i^4\Delta t}{c^3} \lp 1+\frac{t}{t_{\rm sd}} \rp^{-2},
	\label{eq:deltae}
\ee
where $B_*$ is the dipole magnetic field, $R_*$ is the radius, $\Omega_i$ is the initial spin frequency, and the spin down timescale is
\be
	t_{\rm sd} = \frac{Ic^3}{2(B_*R_*^3)^2\Omega_i^2},
\ee
where $I$ is the NS's moment of inertia.

To produce FRBs powered by the NS spin down, $\Delta E$ must be sufficiently large and $t_{\rm sd}$ must be sufficiently long that the emission is not strongly impacted by free-free absorption. In Figure \ref{fig:constraints}, I present the constraints implied by this. Here I use $I=10^{45}\,{\rm g\,cm^2}$, $R_*=10\,{\rm km}$, and $\Delta t=10^{-3}\,{\rm s}$. Lines of constant $\Delta E$ and $t_{\rm sd}$ are labeled. Given the uncertainty in the distance and beaming factor of FRBs, their actual energetics remains highly uncertain. A $1\,{\rm Jy}$ source at $1\,{\rm GHz}$ and a distance of $1\,{\rm Gpc}$ gives an isotropic energy of $\Delta E\approx10^{39}\,{\rm erg}$. Besides just distance uncertainties, the actual energy associated with an FRB could be strongly impacted by beaming effects and/or if the observed radio emission is merely a fraction of the total energy released.

The strongest constraint from Figure \ref{fig:constraints} is that $t_{\rm sd}$ must be sufficiently long that free-free absorption does not dominate as shown in Figure \ref{fig:freefree}. This means that the NS must have a relatively low magnetic field in the range of $B_*\lesssim10^{11}-10^{12}\,{\rm G}$. Although on the face of it one might think that a magnetar strength field of $\approx10^{14}-10^{15}\,{\rm G}$ would be preferable for powering an FRB, instead this would in fact cause the NS to spin down much too fast. Most of the magnetar's energy would come out when the SNR is still optically thick to radio emission. In the future, once the distance scale and energetics of FRBs are better understood, the NS spin can also be constrained. For example, if indeed it was found that $\Delta E\approx 10^{39}\,{\rm erg}$, then $P_i\lesssim1.5\,{\rm ms}$ along with the tight upper limits on $B_*\lesssim10^{11}\,{\rm G}$.

\section{Discussion and Conclusions}
\label{sec:discussion}

I have considered the impact of an SNR on FRBs as a function of the ejecta mass $M_{\rm ej}$ and ISM density $n_0$. The reverse shock generated as ejecta interacts with the ISM heats the SNR, ionizing the material and impacting any potential radio emission coming from within the SNR. This could potentially contribute to the DM of the FRB and add a time dependent component to the DM. These effects are likely difficult to measure because free-free absorption attenuates the radio waves before times of $\approx100-500\,{\rm yrs}$ after the SN. This timescale scales up with increasing $M_{\rm ej}$ and down with increasing $n_0$. Magnetic fields generated at the reverse shock may produce Faraday rotation on the FRB at a level consistent with one observed RM. Depending on $M_{\rm ej}$ and $n_0$, this can occur late enough that it will not be affected by free-free absorption. A wide range of RM values are expected as summarized in the upper panel Figure \ref{fig:freefree}, and this may provide an additional useful probe of the FRB environment.

Going into the future, detecting $\nu_{\rm crit}$ as plotted in the lower panel of Figure \ref{fig:freefree} will be key for constraining the properties of the region around the FRB. The upcoming Canadian Hydrogen Intensity Mapping Experiment \citep[CHIME,][]{chime} will be ideally suited to do this since it will collect a large number of FRBs (estimated at $\sim1-10$ per hour) and be sensitive to a frequency range of $400-800\,{\rm MHz}$ where the free-free absorption is expected to occur. Note though that $\nu_{\rm crit}$ passes through the radio bands of interest relatively quickly in comparison to the full evolution, which may make finding sources with just the right range of ages challenging. On the other hand, if FRBs are never seen at lower frequencies, Figure \ref{fig:freefree} may be used to provide an upper limit to the age of the SNR. Searching for counterparts at higher electromagnetic frequencies may also constrain the presence of an SNR \citep{Murase16,Lyutikov16b}.

The combined requirements of a large amount of spin energy and a long spin down time puts tight constraints on the properties of the NS, as summarized in Figure \ref{fig:constraints}. This begs the question of whether the high rate inferred for FRBs, one of their most outstanding features, can be met by such stringent conditions. Note that these constraints are specific to spin energy and that bursts from a NS tapping its magnetic energy instead are not constrained in the same way. Perhaps this means that some sort of outburst from a magnetar analogous to SGRs is a better candidate \citep[e.g.,][]{Kulkarni14,Kulkarni15,Lyubarsky14}. Better understanding the FRB rate, environment, and how often they repeat will be key questions to address in the near future.

\acknowledgments
I thank the organizers and participants of the UNLV Transients Workshop (April 11 and 12, 2016) where much of this work was inspired. I also thank Liam Connor, Jonathan Katz, and Kohta Murase for feedback on a previous draft, and Carles Badenes and Mark Seibert for helpful discussions.

\bibliographystyle{apj}

\begin{thebibliography}{}
\expandafter\ifx\csname natexlab\endcsname\relax\def\natexlab#1{#1}\fi

\bibitem[{{Bandura} {et~al.}(2014){Bandura}, {Addison}, {Amiri}, {Bond},
  {Campbell-Wilson}, {Connor}, {Cliche}, {Davis}, {Deng}, {Denman}, {Dobbs},
  {Fandino}, {Gibbs}, {Gilbert}, {Halpern}, {Hanna}, {Hincks}, {Hinshaw},
  {H{\"o}fer}, {Klages}, {Landecker}, {Masui}, {Mena Parra}, {Newburgh}, {Pen},
  {Peterson}, {Recnik}, {Shaw}, {Sigurdson}, {Sitwell}, {Smecher}, {Smegal},
  {Vanderlinde}, \& {Wiebe}}]{chime}
{Bandura}, K., {Addison}, G.~E., {Amiri}, M., {et~al.} 2014, in \procspie, Vol.
  9145, Ground-based and Airborne Telescopes V, 914522

\bibitem[{{Connor} {et~al.}(2016){Connor}, {Sievers}, \& {Pen}}]{Connor16}
{Connor}, L., {Sievers}, J., \& {Pen}, U.-L. 2016, \mnras, 458, L19

\bibitem[{{Cordes} \& {Wasserman}(2016)}]{Cordes16}
{Cordes}, J.~M., \& {Wasserman}, I. 2016, \mnras, 457, 232

\bibitem[{{Dai} {et~al.}(2016){Dai}, {Wang}, {Wu}, \& {Huang}}]{Dai16}
{Dai}, Z.~G., {Wang}, J.~S., {Wu}, X.~F., \& {Huang}, Y.~F. 2016, ArXiv
  e-prints, arXiv:1603.08207

\bibitem[{{Falcke} \& {Rezzolla}(2014)}]{Falcke14}
{Falcke}, H., \& {Rezzolla}, L. 2014, \aap, 562, A137

\bibitem[{{Geng} \& {Huang}(2015)}]{Geng15}
{Geng}, J.~J., \& {Huang}, Y.~F. 2015, \apj, 809, 24

\bibitem[{{Hansen} \& {Lyutikov}(2001)}]{Hansen01}
{Hansen}, B.~M.~S., \& {Lyutikov}, M. 2001, \mnras, 322, 695

\bibitem[{{Karastergiou} {et~al.}(2015){Karastergiou}, {Chennamangalam},
  {Armour}, {Williams}, {Mort}, {Dulwich}, {Salvini}, {Magro}, {Roberts},
  {Serylak}, {Doo}, {Bilous}, {Breton}, {Falcke}, {Grie{\ss}meier}, {Hessels},
  {Keane}, {Kondratiev}, {Kramer}, {van Leeuwen}, {Noutsos}, {Os{\l}owski},
  {Sobey}, {Stappers}, \& {Weltevrede}}]{Karastergiou15}
{Karastergiou}, A., {Chennamangalam}, J., {Armour}, W., {et~al.} 2015, \mnras,
  452, 1254

\bibitem[{{Kashiyama} {et~al.}(2013){Kashiyama}, {Ioka}, \&
  {M{\'e}sz{\'a}ros}}]{Kashiyama13}
{Kashiyama}, K., {Ioka}, K., \& {M{\'e}sz{\'a}ros}, P. 2013, \apjl, 776, L39

\bibitem[{{Katz}(2015)}]{Katz15}
{Katz}, J.~I. 2015, ArXiv e-prints, arXiv:1512.04503

\bibitem[{{Katz}(2016)}]{Katz16}
---. 2016, ArXiv e-prints, arXiv:1604.01799

\bibitem[{{Keane} {et~al.}(2012){Keane}, {Stappers}, {Kramer}, \&
  {Lyne}}]{Keane12}
{Keane}, E.~F., {Stappers}, B.~W., {Kramer}, M., \& {Lyne}, A.~G. 2012, \mnras,
  425, L71

\bibitem[{{Kulkarni} {et~al.}(2015){Kulkarni}, {Ofek}, \& {Neill}}]{Kulkarni15}
{Kulkarni}, S.~R., {Ofek}, E.~O., \& {Neill}, J.~D. 2015, ArXiv e-prints,
  arXiv:1511.09137

\bibitem[{{Kulkarni} {et~al.}(2014){Kulkarni}, {Ofek}, {Neill}, {Zheng}, \&
  {Juric}}]{Kulkarni14}
{Kulkarni}, S.~R., {Ofek}, E.~O., {Neill}, J.~D., {Zheng}, Z., \& {Juric}, M.
  2014, \apj, 797, 70

\bibitem[{{Loeb} {et~al.}(2014){Loeb}, {Shvartzvald}, \& {Maoz}}]{Loeb14}
{Loeb}, A., {Shvartzvald}, Y., \& {Maoz}, D. 2014, \mnras, 439, L46

\bibitem[{{Lorimer} {et~al.}(2007){Lorimer}, {Bailes}, {McLaughlin},
  {Narkevic}, \& {Crawford}}]{Lorimer07}
{Lorimer}, D.~R., {Bailes}, M., {McLaughlin}, M.~A., {Narkevic}, D.~J., \&
  {Crawford}, F. 2007, Science, 318, 777

\bibitem[{{Luan} \& {Goldreich}(2014)}]{Luan14}
{Luan}, J., \& {Goldreich}, P. 2014, \apjl, 785, L26

\bibitem[{{Lyubarsky}(2014)}]{Lyubarsky14}
{Lyubarsky}, Y. 2014, \mnras, 442, L9

\bibitem[{{Lyutikov} {et~al.}(2016){Lyutikov}, {Burzawa}, \&
  {Popov}}]{Lyutikov16}
{Lyutikov}, M., {Burzawa}, L., \& {Popov}, S.~B. 2016, ArXiv e-prints,
  arXiv:1603.02891

\bibitem[{{Lyutikov} \& {Lorimer}(2016)}]{Lyutikov16b}
{Lyutikov}, M., \& {Lorimer}, D.~R. 2016, ArXiv e-prints, arXiv:1605.01468

\bibitem[{{Masui} {et~al.}(2015){Masui}, {Lin}, {Sievers}, {Anderson}, {Chang},
  {Chen}, {Ganguly}, {Jarvis}, {Kuo}, {Li}, {Liao}, {McLaughlin}, {Pen},
  {Peterson}, {Roman}, {Timbie}, {Voytek}, \& {Yadav}}]{Masui15}
{Masui}, K., {Lin}, H.-H., {Sievers}, J., {et~al.} 2015, \nat, 528, 523

\bibitem[{{McKee} \& {Truelove}(1995)}]{McKee95}
{McKee}, C.~F., \& {Truelove}, J.~K. 1995, \physrep, 256, 157

\bibitem[{{Murase} {et~al.}(2016){Murase}, {Kashiyama}, \&
  {Meszaros}}]{Murase16}
{Murase}, K., {Kashiyama}, K., \& {Meszaros}, P. 2016, ArXiv e-prints,
  arXiv:1603.08875

\bibitem[{{Pen} \& {Connor}(2015)}]{Pen15}
{Pen}, U.-L., \& {Connor}, L. 2015, \apj, 807, 179

\bibitem[{{Piro}(2012)}]{Piro12}
{Piro}, A.~L. 2012, \apj, 755, 80

\bibitem[{{Rane} {et~al.}(2016){Rane}, {Lorimer}, {Bates}, {McMann},
  {McLaughlin}, \& {Rajwade}}]{Rane16}
{Rane}, A., {Lorimer}, D.~R., {Bates}, S.~D., {et~al.} 2016, \mnras, 455, 2207

\bibitem[{{Ravi} {et~al.}(2015){Ravi}, {Shannon}, \& {Jameson}}]{Ravi15}
{Ravi}, V., {Shannon}, R.~M., \& {Jameson}, A. 2015, \apjl, 799, L5

\bibitem[{{Rowlinson} {et~al.}(2016){Rowlinson}, {Bell}, {Murphy}, {Trott},
  {Hurley-Walker}, {Johnston}, {Tingay}, {Kaplan}, {Carbone}, {Hancock},
  {Feng}, {Offringa}, {Bernardi}, {Bowman}, {Briggs}, {Cappallo}, {Deshpande},
  {Gaensler}, {Greenhill}, {Hazelton}, {Johnston-Hollitt}, {Lonsdale},
  {McWhirter}, {Mitchell}, {Morales}, {Morgan}, {Oberoi}, {Ord}, {Prabu},
  {Udaya Shankar}, {Srivani}, {Subrahmanyan}, {Wayth}, {Webster}, {Williams},
  \& {Williams}}]{Rowlinson16}
{Rowlinson}, A., {Bell}, M.~E., {Murphy}, T., {et~al.} 2016, \mnras, 458, 3506

\bibitem[{{Rybicki} \& {Lightman}(1979)}]{Rybicki79}
{Rybicki}, G.~B., \& {Lightman}, A.~P. 1979, {Radiative processes in
  astrophysics}

\bibitem[{{Scholz} {et~al.}(2016){Scholz}, {Spitler}, {Hessels}, {Chatterjee},
  {Cordes}, {Kaspi}, {Wharton}, {Bassa}, {Bogdanov}, {Camilo}, {Crawford},
  {Deneva}, {van Leeuwen}, {Lynch}, {Madsen}, {McLaughlin}, {Mickaliger},
  {Parent}, {Patel}, {Ransom}, {Seymour}, {Stairs}, {Stappers}, \&
  {Tendulkar}}]{Scholz16}
{Scholz}, P., {Spitler}, L.~G., {Hessels}, J.~W.~T., {et~al.} 2016, ArXiv
  e-prints, arXiv:1603.08880

\bibitem[{{Spitler} {et~al.}(2014){Spitler}, {Cordes}, {Hessels}, {Lorimer},
  {McLaughlin}, {Chatterjee}, {Crawford}, {Deneva}, {Kaspi}, {Wharton},
  {Allen}, {Bogdanov}, {Brazier}, {Camilo}, {Freire}, {Jenet},
  {Karako-Argaman}, {Knispel}, {Lazarus}, {Lee}, {van Leeuwen}, {Lynch},
  {Ransom}, {Scholz}, {Siemens}, {Stairs}, {Stovall}, {Swiggum},
  {Venkataraman}, {Zhu}, {Aulbert}, \& {Fehrmann}}]{Spitler14}
{Spitler}, L.~G., {Cordes}, J.~M., {Hessels}, J.~W.~T., {et~al.} 2014, \apj,
  790, 101

\bibitem[{{Spitler} {et~al.}(2016){Spitler}, {Scholz}, {Hessels}, {Bogdanov},
  {Brazier}, {Camilo}, {Chatterjee}, {Cordes}, {Crawford}, {Deneva}, {Ferdman},
  {Freire}, {Kaspi}, {Lazarus}, {Lynch}, {Madsen}, {McLaughlin}, {Patel},
  {Ransom}, {Seymour}, {Stairs}, {Stappers}, {van Leeuwen}, \&
  {Zhu}}]{Spitler16}
{Spitler}, L.~G., {Scholz}, P., {Hessels}, J.~W.~T., {et~al.} 2016, \nat, 531,
  202

\bibitem[{{Thornton} {et~al.}(2013){Thornton}, {Stappers}, {Bailes},
  {Barsdell}, {Bates}, {Bhat}, {Burgay}, {Burke-Spolaor}, {Champion}, {Coster},
  {D'Amico}, {Jameson}, {Johnston}, {Keith}, {Kramer}, {Levin}, {Milia}, {Ng},
  {Possenti}, \& {van Straten}}]{Thornton13}
{Thornton}, D., {Stappers}, B., {Bailes}, M., {et~al.} 2013, Science, 341, 53

\bibitem[{{Wang} {et~al.}(2016){Wang}, {Yang}, {Wu}, {Dai}, \& {Wang}}]{Wang16}
{Wang}, J.-S., {Yang}, Y.-P., {Wu}, X.-F., {Dai}, Z.-G., \& {Wang}, F.-Y. 2016,
  ArXiv e-prints, arXiv:1603.02014

\bibitem[{{Zhang}(2016)}]{Zhang16}
{Zhang}, B. 2016, ArXiv e-prints, arXiv:1602.04542

\end{thebibliography}

\end{document}